# Analysis of Intelligent Classifiers and Enhancing the Detection Accuracy for Intrusion Detection System

Mohanad Albayati and Biju Issac
*School of Computing, Teesside University,
Middlesbrough, England, UK
E-mail: b.issac@tees.ac.uk*

**Abstract**

In this paper we discuss and analyze some of the intelligent classifiers which allows for automatic detection and classification of networks attacks for any intrusion detection system. We will proceed initially with their analysis using the WEKA software to work with the classifiers on a well-known IDS (Intrusion Detection Systems) dataset like NSL-KDD dataset. The NSL-KDD dataset of network attacks was created in a military network by MIT Lincoln Labs. Then we will discuss and experiment some of the hybrid AI (Artificial Intelligence) classifiers that can be used for IDS, and finally we developed a Java software with three most efficient classifiers and compared it with other options. The outputs would show the detection accuracy and efficiency of the single and combined classifiers used.

## 1. Introduction

The computer networks expand on a daily basis and the users of Internet are increasing. The sharing of information is turning the world into a small village. The technology of exchanging information across networks had improved the efficiency of data transfer, but it also has made more opportunity for cyber-attacks. All of these possible network attacks make users, organizations and government agencies to want to protect their systems from intrusions. The intrusion can be defined as the ability to break through a system and trying to compromise its integrity, availability, confidentiality or quality of service (Abraham and Patra, 2012). There are different defense measures employed by most organizations to prevent the computer networks and sensitive data from intrusion or attacks like authentication, firewalls and physical security. All of these measures are good but they do not protect against sophisticated attacks - say like buffer overflow attacks which makes use of the weakness in an application and cause enormous security threat. That's when the need for an intrusion detection system (IDS) began to appear. They are like a second line of security defense. According to (Abraham and Patra, 2012) the IDS can be defined as a system of observing suspicious actions that happens on computer networks to detect users who are not permitted access, trying to breach network devices. There are two typical methods of IDS that can be implemented on computer networks, namely - Signature based and Anomaly based (Benferhat and Tabia, 2004), and there are some that is a mix between those two (Elvis, 2004). Signature based detection uses a signature database to detect suspicious activity, where each signature represent a print of known attack. These systems are only as good as their database. Therefore the database need to be updated continuously to ensure there is information about the new intrusions. Anomaly based intrusion detection system builds a profile of normal system behaviour and detects any deviation from the profile to identify possible attacks. The profile can be constructed using machine learning techniques and data mining and should be upgraded regularly. The advantage of anomaly based over signature based is that it can identify non-trivial attacks but it has a high tendency for generating false alarms (Aydin, 2009). The main problems with current IDS are efficiency and accuracy in detecting intrusions (Hofmann and Sick, 2011). In this paper we are planning to study and analyse the different intelligence classifiers of IDS, test some of the hybrid approaches, and design a hybrid software system that can be accurate and efficient at the same time.

This paper is organised as follows. Section 2 is the related works on intrusion detection done, section 3 is introduction to IDS, section 4 is the discussion on dataset used, section 5 is the analysis of IDS classifiers, section 6 is on the developed software system, section 7 is the discussion and limitation of our work and section 8 is the conclusion.

## 2. Related Works

The first concept of intrusion detection was introduced through a paper by (James P. Anderson, 1980) where the authors introduced a model that develops a security monitoring surveillance to detect anomalies in user behaviour.

(Lee and Stolfo, 1998) proposed a systematic framework that employs data mining techniques to detect intrusions. (Schultz, Zadok, Stolfo, 2001) proposed a framework which uses data mining classifiers to train multiple classifiers on a set of malicious and benign executables to detect new examples. (Nadiammai and Hemalatha, 2012) presented a study of all ruled based classifiers to predict their effectiveness based on accuracy, time, specificity, sensitivity and error.

(Hwang, Lee, and Lee, 2007) proposed a three-tier architecture of IDS which consist of three lists - black list, white list and multi-class. The black list contain any known attacks from the traffic, the white list contain the rest of the normal traffic, and the third list called multi-class contains anomalies that are detected in the normal traffic. (Tavallaee, Bagheri, Wei, and Ghorbani, 2009) presented a study of each feature in KDD '99 intrusion detection dataset. (Subramanian, Srinivasan, and Ramasa, 2012) aimed to classify NSL-KDD dataset using Random Tree classifier with respect to their metric data and study their performance.

(Lippmann, Haines, Fried, J. Korba, and Das, 2000) presented a comparison study to the various data mining classification techniques for intrusion detection. (Srinivasulu, Nagaraju, Kumar, and Rao, 2009) presented different data mining techniques named CART, Naive Bayesian, and artificial neural network and evaluated the performance of each techniques using a confusion matrix. (Kalyani and Lakshmi, 2012) presented a comparison study between the techniques such as Naive Bayes, J48, OneR, PART, and RBF network classifier using NSL-KDD dataset; They also discussed the advantages of using NSL-KDD dataset over KDDCUP'99. (Reddy, IAENG, Reddy, and Rajulu, 2011) presented a survey of various techniques and the enhancement of IDS. (Neethu, 2012) explained about the IDS framework that is a combination of Naïve Bayes and Principal Component Analysis classifier that helped to increase the speed of performance.

## 3. Introduction to IDS

Intrusion detection systems or IDS became very important to office security now-a-days. However many experts are still unsure about the function of these systems, as to why we use them and how they perform. The good number of security threats come from inside the organization networks because of authorized indignant employees. Or sometimes the attacks can be through someone with stolen credentials of a valid employee, which can be very difficult to trace. The other attacks could come from outside users through denial of service attacks or through hack attempts to penetrate the network. Intrusion detection systems are the only means to detect those attacks and respond to threats that occur from both inside and outside the organizational network. Intrusion detection systems are necessary for a complete security infrastructure. (BAC, 1999) said that using IDS allows you to completely supervise a network, regardless of the action being taken, and that information will always exist to determine the nature of the security threat and its source. Today most medium size organizations have installed some form of intrusion detection or something similar. Network attacks and intrusion is motivated by financial, political, military or personal reasons and every network is a potential target. So owners of any official network should consider some form of IDS, the networks are always at risk of attacks.

In early 2014 the cyber-attacks had caused security breach of eBay employee log-ins, allowing access to the contact and log-in information of around 233 million eBay users. The Yahoo e-mail service for 273 million users was hacked in early 2014, although the exact number of accounts affected was not disclosed. In 2013 the Facebook was attacked by hackers who exploited a previously unknown loophole in its computer system. In the same year the Facebook hackers attacked Apple computers, though no data appeared to have been stolen and Burger King's twitter account became victim to hackers as well as it began sending out pro-McDonald's message. In 2011 Sony was attacked and hackers stole private details of more than a million users. In 2007, TJX, the parent company of discount stores T.J. Maxx and Marshalls, disclosed that thieves had stolen data of tens of millions of credit and debit cards. There was a reported denial of service attacks in 2000 against Amazon and E-bay. These consistent and recent attacks shows the need to have an intrusion detection system especially for commercial network and websites.

Intrusion detection is the process of monitoring networks and computers for unauthorized access, suspicious activity or file modification. IDS can also monitor network traffic to detect if the system is being targeted by network attacks like the different types of denial of service attacks. The two types of intrusion detection are Host-Based (HIDS) and Network-Based (NIDS) approaches. Each of these attacks has different ways to monitor. HIDS examine the personal data held on computers, while NIDS looks at the exchange of data between computers.

**3.1. *IDS Approaches and Techniques***

For each of the two types of intrusion techniques - HIDS and NIDS, there are four basic techniques to detect the attacks - Anomaly detection, Misuse detection, Target monitoring and Stealth probes.

3.1.1. *Anomaly Detection*

Like the name suggest, anomaly detection is searching for suspicious behaviour that the user doesn't normally perform. Example of suspicious behaviour can be as follows: the user log in more than 20 times a day, or accessing e-mail that they are not allowed to, or log in at 2 am or out of the office hours etc. This will be considered as an unusual behaviour and will alert the system administrator.

3.1.2. *Misuse Detection*

Misuse detection or signature detection is used to identify a specific known pattern of unauthorized behaviour to predict similar attempts. These patterns are called signatures. For example an improper FTP, depending on the seriousness of the signature and alarm could be triggered or a notification could be sent to the admin to handle it.

3.1.3. *Target Monitoring*

These systems do not monitor behaviour or look for signatures; they only look for modification in specific files and they are designed to undercover the unauthorized modification after it occurs. They can be checked by computer through cryptographic hash on files beforehand and compare the old files with new files. These systems can be easy implemented and doesn't require constant monitoring by the administrator.

3.1.4. *Stealth probes*

This approach attempts to detect any attacks that is carried out for prolonged periods of time. For example the attacks will check for system vulnerabilities and open ports and collect data and information about the system and then launch the attack say, after two months of the original system infection. This method combines anomaly detection and misuse detection to discover suspicious behaviour.

**4. Discussion on Dataset Used**

We wanted to discuss on the details of dataset used for our experiments. This would help to see what kind of network attacks are addressed in our work.

**4.1. *NSS-KDD Dataset***

The DARPA Intrusion Detection Evaluation Program by MIT Lincoln Labs in 1998 wanted to research into intrusion detection. A wide variety of intrusions simulated in a military network was generated and that became the 1999 KDD intrusion detection dataset. This data contained nine weeks of raw TCP dump data for a simulated U.S. Air Force LAN with a number of network attacks. The attacks fall into four main categories: (1) DoS – Denial of service (2) U2R - Unauthorized access from a remote machine (3) R2L - Unauthorized access to local super-user privileges (4) Probe - Surveillance and other probing. DoS attack are designed to consume all network bandwidth and will look like normal traffic. The user to root (U2R) attack happens on a local machine to elevate the user privileges to that of the super user. Remote to local (R2L) activity are attempts to login to a computer or device from outside. Probe activity is done over the network to collect the details of devices on the network.

The KDD training dataset consisted of 494,019 records where 97,277 (19.69%) were classified as 'normal', 391,458 (79.24%) as DoS, 4,107 (0.83%) as Probe, 1,126 (0.23%) as R2L and 52 (0.01%) as U2R attacks. Each record has 41 attributes described different features and a label was assigned to each either as an 'attack' type or as 'normal' type. (Siddiqui and Naahid, 2013), (KDD Cup 1999 Data, 2014). Because of this labelling we did not need to do any tuning to the dataset.

The basic features of individual TCP connections contained the following features: length (number of seconds) of the connection, type of the protocol like tcp or udp, network service on the destination like http or telnet, number of data bytes from source to destination, number of data bytes from destination to source, normal or error status of the connection, 1 if connection is from/to the same host/port; 0 otherwise, number of wrong fragments and number of urgent packets. The content features within a connection suggested by domain knowledge contained the following features: number of 'hot' indicators, number of failed login attempts, 1 if successfully logged in; 0 otherwise, number of 'compromised' conditions, 1 if root shell is obtained and 0 otherwise, 1 if 'su root' command attempted and 0 otherwise, number of 'root' accesses, number of file creation operations, number of shell prompts, number of operations on access control files, number of outbound commands in an ftp session, 1 if the login belongs to the 'hot' list and 0 otherwise, 1 if the login is a 'guest' login and 0 otherwise (KDD Cup 1999 Data, 2014). Table 1 shows the attack dataset showing the type of attacks grouped as four categories.

Table 1. Type of attacks grouped as four categories.

| Attacks in Dataset | Attack Type |
| --- | --- |
| DoS | apache2, smurf, neptune, dosnuke, land, pod, back, teardrop, tcpreset, syslogd, crashiis, arppoison, mailbomb, selfping, processtable, udpstorm, warezclient |
| Probe | portsweep, ipsweep, queso, satan, msscan, ntinfoscan, lsdomain, illegal-sniffer |
| R2L | dict, netcat, sendmail, imap, ncftp, xlock, xsnoop, sshtrojan, framespoof, ppmacro, guest, netbus, snmpget, ftpwrite, httptunnel, phf, named |
| U2R | sechole, xterm, eject, ps, nukepw, secret, perl, yaga, fdformat, ffbconfig, casesen, ntfsdos, ppmacro, loadmodule, sqlattack |

The NSL-KDD is a dataset for intrusion detections systems and it is originally from KDD'99 dataset but it fixes some of the inheritance problems that are mentioned in (Tavallaee, Bagheri, Lu, and Ghorbani, 2009). Although the NSL-KDD dataset still have some problems, it is still good for training the IDS and it has a reasonable amount of records. This advantage will make it perfect for running experiments and the evaluation of records will be consistent and comparable.

The NSL-KDD dataset has the following advantages over the original KDD dataset. As it avoids the duplicate records in the training set and in test sets, there will be no bias in the classifiers towards records that are more frequent during training and the performance of the learners are not biased by the classifiers that have better detection rates on the frequent records during testing. The number of selected records from each difficulty level group is inversely proportional to the percentage of records in the original KDD dataset. Thus there is a wiser range in the classification rates of distinct machine learning methods, which allows an accurate evaluation of different learning classifiers (NSL-KDD, 2014).

### 4.2. *WEKA Software Study*

The WEKA (2003) software is a program written in Java to test out the different available artificial intelligence (AI) classifiers. After studying the software, we started to test the different classifiers. This software is a really helpful tool to decide which classifiers gives the best results, after testing it on WEKA software using NSL-KDD dataset. Like mentioned previously, the NSL-KDD is a dataset which is better than the original KDD'99 dataset and is a good baseline dataset to compare different intrusion detection methods. The best results given was for Random Forest (RF) with 99.89% accuracy, followed by Random Tree (RT) with 99.77% accuracy and Naïve Bayes (NB) with 90.38% accuracy.

### 4.3. *Arguments for and against NSL-KDD dataset*

Thomas and Sharma et al. (2008) states the usefulness of DARPA dataset for IDS evaluation. The DARPA evaluation dataset has been found to have the required potential in modelling the attacks that appear commonly on the network traffic. They affirm that the dataset can be considered as the base line of any research. The paper concludes that it can be used to evaluate the IDSs in the present scenario, against the notion that it is a very outdated dataset, unable to accommodate the latest trend in attacks. Tavallaee and Bagheri et al. (2009) argue that although the KDD Cup '99 datasets suffer from various problems, they are still an effective benchmark to compare different intrusion detection methods. To address some of the known issues a revised version of the datasets called NSL-KDD was created. We felt that the analysis of NSL-KDD will yield a predictable performance results for the intrusion detection algorithms we are using.

There are some arguments against using this dataset. McHugh (2000) wrote a detailed critique identifying shortcomings of KDD dataset evaluations where he claimed that the evaluation failed to verify that the network realistically simulated a real-world network. Mahoney and Chan (2003) also found problems as they looked at the content of the 1999 DARPA evaluation tcpdump data. They found that the simulated traffic contains irregularities where many of the network attributes with large range in real-world traffic, have a small and fixed range in the simulation.

## 5. Analysis of Intrusion Detection System Classifiers

In this section we explain the details of experiments done with different classifiers and the results achieved.

### 5.1. *Experiments Performed*

Several experiments were performed to test out the best performance of each of the three selected classifiers – Naïve Bayes, Random Tree and Random Forest. All experiments were conducted on VAIO Laptop with Intel(R) Core I (3), 2.53 GHz CPU and 4.00GB RAM with 250GB HDD. There were a total of 10 experiments for each of these classifiers as listed below.

#### 5.1.1. *Naive Bayes*

Naive Bayes classifier is group of simple classifiers using Bayes' probability theorem with strong independence assumptions between the features of what is being binary classified (with two states – yes or no). This experiment was performed using WEKA software on NSL-KDD dataset, the classifier used was Naïve Bayes and the test option used was - cross validation of 10 cross folds. From table 1 it is evident that the intrusion detection rate is 90.38% with alarm rate of 9.62%. It is error prone with root mean square value of 0.3058 which means it performs poorly compared to other classifiers.

#### 5.1.2. *Applying Discretize filter to Naïve Bayes*

We tried applying discretize filter to Naïve Bayes. Discretization uses a set of predefined intervals and grouping the featured values according to those interval values. Or in other words, discretization involves dividing an attribute's values into a number of intervals so that each interval can be treated as one value of a discrete attribute. Thus the learning complexity of the Naïve Bayes classifier can be reduced. The experiment was done as before. As in table 1, you can notice the change in accuracy after applying the filter have gone significantly up from 90% to over 97%, and the build time took only 0.12 seconds while before it was 1.57 seconds. You can also notice a lower false detection rate of 2.87% while it was 9.62% before, and that shows the filter is getting much higher results than the normal Naive Bayes classifier.

### 5.1.3. *Random Tree*

The experiment was done as before, but with Random Tree (RT). RT used a certain number of randomly chosen attributes at each node of a decision tree. It is a predictive model that uses a set of binary rules and can be used for classification or regression applications. It is quite easy to interpret the decision rules. The classification is quick once the rules are designed. From table 1 we can infer that Random Tree intrusion detection is quite high with 99.77% accuracy with extremely low false alarm rate of 0.11%, which is an excellent performance. It is slower than Naïve Bayes where the model build took 2.59 seconds. A high F-Measure of 99% can also be observed.

### 5.1.4. *Random Forest*

Again the experiment was done as before, but with Random Forest (RF). RF is an ensemble classifier that combines the results from different models using many Random Tree models. Here there is no need to prune trees and overfitting is not a problem. As seen from table 1, it is evident that Random Forest intrusion detection rate is high with 99.89% accuracy with extremely low false alarm rate of 0.11%, which is a very high performance. It is slower than Naïve Bayes with model build that took 22.33 seconds. But a high F-Measure of 99% can be noted.

### 5.1.5. *Comparing the classifiers performance using ROC curve*

The "Receiver Operating Characteristic" (ROC) curve is an alternative to accuracy for the evaluation of learning classifiers on natural datasets. The curve is plotted by using the true positive rate against the false positive rate at various threshold settings. We tried to compare the three classifier's performance that we worked on - Naïve Bayes, Random Tree and Random Forest. The smaller the ROC curve and the more close it is to value 1 on y-axis the better the performance of the classifier. Refer to Figure 1. Naïve Bayes performance was slightly less good than Random Tree and Random Forest as we can see there is some curve on the thick line. Random Tree and Random Forest performance was excellent with the lack of curve that indicates a high performance of the classifiers.

Table 1. Performance of AI classifiers (Cross Validation of 10 Cross Folds)

| Parameters | Naïve Bayes | Naïve Bayes with Discretize filter | Random Forest | Random Tree |
|---|---|---|---|---|
| Correctly Classified Instances | 113858 (90.38%) | 122353 (97.13%) | 125835 (99.89%) | 125678 (99.77%) |
| Incorrectly Classified Instances | 12115 (9.62%) | 3620 (2.87%) | 138 (0.11%) | 295 (0.11%) |
| Total Number of Instances | 125973 | 125973 | 125973 | 125973 |
| Root mean squared error | 0.3058 | 0.1612 | 0.0313 | 0.0479 |
| Model Building Time | 1.57 seconds | 0.12 seconds | 22.33 seconds | 2.59 seconds |
| TP Rate | 0.904 | 0.971 | 0.999 | 0.998 |
| FP Rate | 0.101 | 0.032 | 0.001 | 0.002 |
| Recall | 0.904 | 0.971 | 0.999 | 0.998 |
| F-Measure | 0.966 | 0.997 | 0.999 | 0.998 |

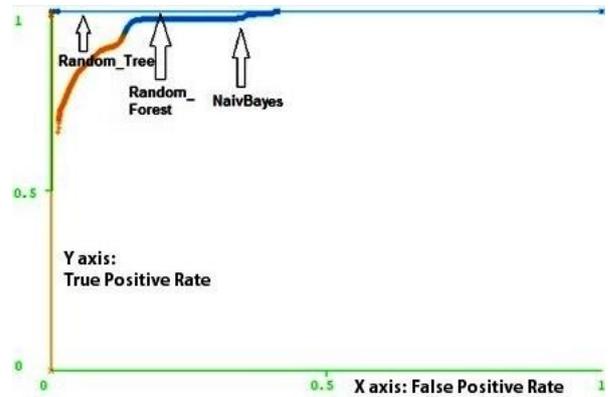

Fig. 1. ROC for - Naïve Bayes, Random Tree and Random Forest (Singular versions)

### 5.1.6. *Filter Method Naïve Bayes and Wrapper Method Naïve Bayes*

To explore some more options under Naïve Bayes as it has lower model building times, we worked on the filter and the wrapper methods. The Filter Method Naïve Bayes uses an attribute evaluator and a ranker to rank the entire features in the dataset. The number of features we want to select from the vector can be defined. Then we can omit the features one at a time that have the lower rank and we can see a predictive accuracy of the classifier. We can only omit a certain number of features until we reach the global minimum – "the point where you cannot omit features". If we omit more than the number of global minimum the dataset will start overfeeding and we will get an increased number of incorrectly classified instances. We ran the ranker with the global minimum of 41, which means we can omit the entire feature from bottom until we reach 41. While omitting and retesting we noticed an increase of accuracy each time as in table 3. The Naïve Bayes accuracy was initially 90.38%, but with the filter the accuracy has gone up to 90.72%.

In the wrapper method we used a subset evaluator and this created all possible subsets from the featured vector. After using the classifier like Naïve Bayes to induce classifiers from the features in each subset, it will then consider the subset of features with which the classification classifier perform the best. We ran the test and the best featured subset was number (3, 4, 17). After elimination of all except for these three, the results were 96.22 % accurate as in table 3. We observed that the detection accuracy was still lower than the best ones so far.

Table 3. Performance of Filter Method Naïve Bayes and Wrapper Method Naïve Bayes (Cross Validation of 10 Cross Folds)

| Parameters | Naïve Bayes – Filter Method | Naïve Bayes – Wrapper Method |
|---|---|---|
| Correctly Classified Instances | 114283 (90.72%) | 121216 (96.22%) |
| Incorrectly Classified Instances | 11690 (9.28%) | 4757 (3.78%) |
| Total Number of Instances | 125973 | 125973 |
| Root mean squared error | 0.3007 | 0.193 |
| Model Building Time | 2.19 seconds | 1.62 seconds |
| TP Rate | 0.907 | 0.962 |
| FP Rate | 0.1 | 0.038 |
| Recall | 0.907 | 0.962 |
| F-Measure | 0.968 | 0.984 |

### 5.1.7. *Combining Three Classifiers – The Best Accuracy*

After performing all of previous experiments we combined the three classification classifiers - Naïve Bayes (discretized), Random Tree, Random Forest on Weka, and we compared their performance in ROC curve. The result was high performance with 99.9% accuracy. So we decided to use these three classifiers to build a software system to detect intrusions. Refer to figure 2 for ROC curve in comparison to figure 1. There is no curve at all on the thick line. As stated before the smaller the ROC curve and the more close it is to value 1 on y-axis the better the performance of the classifier.

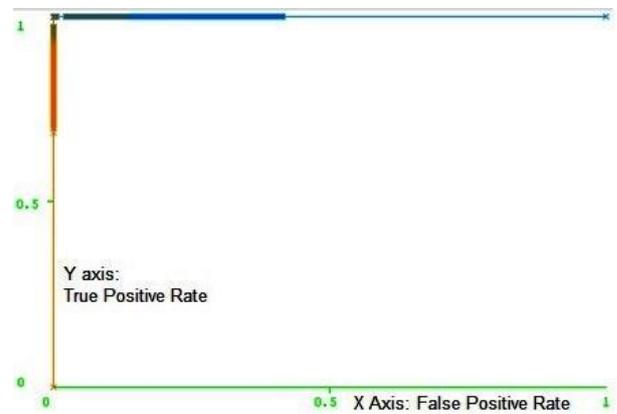

Fig. 2. ROC for - Naïve Bayes, Random Tree and Random Forest (Combined version)

### 6. Developed Software System

In order to test these classifiers and their performance we developed a software in Java to detect intrusions on a network or on a dataset. Using software libraries and Java compiler, this system will function by first training discretized Naïve Bayes classifier separately using K2 learning process. The reason we choose K2 is because it shows high performance, and it can improve the intrusion detection of Naïve Bayes classifier. After training the Naïve Bayes the dataset will go through two other training sessions using Random Tree and Random Forest. These two classifiers will maximize the chance of detecting more intrusions that can pass through Naïve Bayes classifier. After that we will create a method called Junction Tree inference. The idea of this procedure is to construct a data structure called a junction tree which can be used to calculate any query through the message passing on the tree (Jemili, Zaghdoud and Ben Ahmed, 2007).

**6.1.** *K2 Learning Process*

K2 classifier works by finding the best structure amounts to pick the best parents for each node supposing we already know a total ordering on the nodes (Cooper, Herskovits, 1992). K2 is a greedy search classifier and it works as follows. Suppose we already know the ordering of each node, the classifier will incrementally add a set of parents and that addition increases the score of the resulting structure. When no addition of a single parent can increase the score, the classifier will stop adding parents to that node. Based on the assumption that we can add a parent to each node independently, in our system we used this classifier to train our classifier using Bayesian Network which uses Naïve Bayes classifier.

**6.2.** *Naive Bayes*

A Naive Bayes classifier works on the principle that the presence or absence of a specific feature of a class is independent or unrelated to the presence or absence of any other feature.

As per (Statsoft, 2014), be it continuous or categorical - Naive Bayes classifiers can handle a random number of independent variables. Given a set of variables, X = {x1, x2, x..., xn}, we want to construct the posterior probability for the event Cj among a set of possible outcomes C = {c1, c2, c..., cn}. Thus X is the predictors and C is the set of categorical levels present in the dependent variable. As we use the Bayes' rule, we get the following equation (1):

$$p(C_j \mid x_1, x_2, \ldots, x_d) \propto p(x_1, x_2, \ldots, x_d \mid C_j) p(C_j)$$

where p(Cj | x1,x2,x...,xn) is the posterior probability of class membership, i.e., the probability that X belongs to Cj. With the assumption that the conditional probabilities of the independent variables are statistically independent we can decompose the likelihood to a product of terms as in equation (2):

$$p(X \mid C_j) \propto \prod_{k=1}^{d} p(x_k \mid C_j)$$

**6.3.** *Bayesian Network*

The Bayesian network is a representation suited to looking for relationships among a large number of variables. With large set of variables, it is a graphical model that efficiently models the joint probability distribution. It is a graphical representation among a set of random variables (Pearl, 1988). Consider this example as given in Bayesnet.com: Consider the finite set X={X1,…,Xn} of discrete random variables, were each Xi may take the value from a finite set, denoted by Val(Xi). Bayesian network is a graphical representation that encodes joint probability distribution over X. The nodes of the graph correspond to the random variables X1,…,Xn. The graphical links correspond to the direct influence from one variable to another. If there is a direct link between the variable Xi and the Variable Xj then the variable Xi will be a parent to the variable Xj. Figure 3 is an example of Bayesian network (Cooper, 1999).

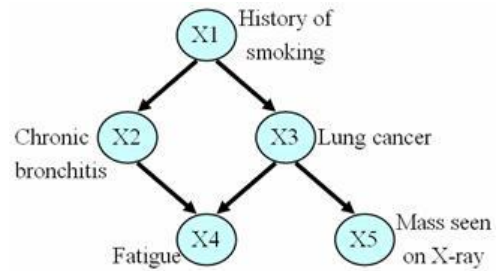

Fig. 3 Bayesian Network (Cooper, 1999).

**6.4.** *Random Tree*

The Decision tree consists of nodes that form a rooted tree. It is a directed tree root node that has no incoming edges but only outgoing ones. Like a binary tree, all other nodes have exactly one incoming edge. A node with outgoing edges is called an internal or test node. All other nodes are called leaves (also known as terminal or decision nodes). In a decision tree, based on a function with input value of attributes, each test or internal node splits the instance space into multiple sub-spaces (Oded, and Lior. 2010) as in figure 4. A Random tree considers K randomly chosen attributes at each node of a decision tree.

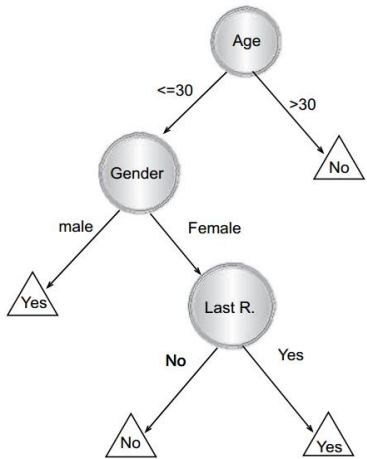

Fig. 4. Decision Tree on Responses to Direct Mailing (Oded, and Lior. 2010)

### 6.5. *Random Forest*

Random forest grows many classification trees. The ideas is as follows. To classify new object from an input vector, put the input vector of each tree in the forest. Each tree will give a classification, and the tree vote for that class. The forest chooses the classification having the most votes. The reason we choose this classifier is because we are using a large dataset, and the trees tend to give high performance when using large datasets.

If the number of cases in the training set is N, sample 'n' cases at random from the original data, where n < N. This sample will be the training set for growing the tree. If there are R input variables, a number r < R is specified at each node, r variables are selected at random out of the R and the best split the node. The value of r will be held constant during the period of tree growing. In random forest there is no pruning, so each tree will grow the largest extent possible. The Random forest as in figure 5 combines trees and though the trees are weak learners, the Random forest is a strong learner. The Random Forest error rate depends on two things: (1) the connections between any two trees in the forest and excess connections in the forest increase the error rate; (2) the strength of each tree in a Forest and increasing the strength of individual trees decreases the error rate.

Most important features of the Random forest are as follows: The accuracy is unpredictable depending on the training set. On large datasets the performance is efficient. It can handle thousands of inputs without having to delete any variables. It can give an estimation of the most important variables to the classifier. It is the most effective method for estimating missing data and maintain accuracy when a large proportion of data is missing. It can balance errors in class population for unbalanced datasets. The generated forests can be saved for future uses on other data. The computed prototypes

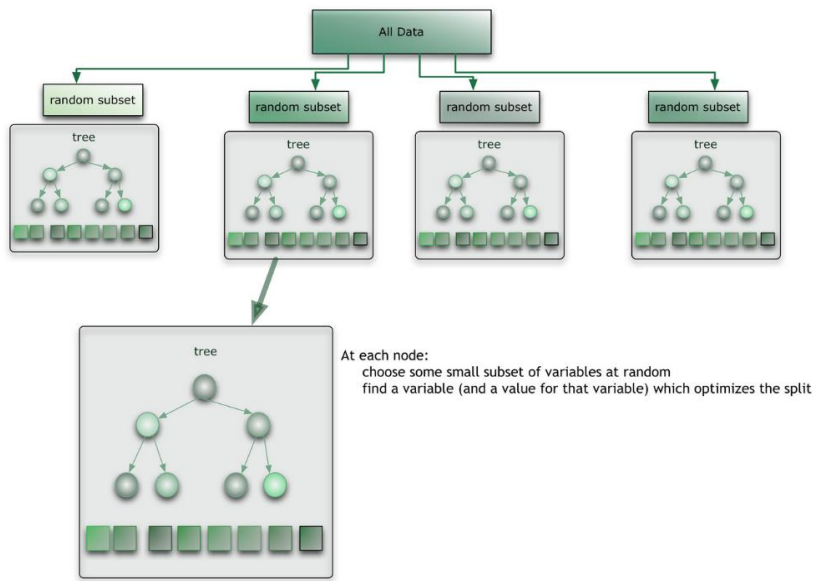

Fig. 5. Random Forest (CitizenNet and Blackwell, A., 2012).

can give information about the relations between variables. It can detect variables interactions.

### 6.6. *Implemented Software Building Blocks*

The software system that we developed in Java is shown in figure 6 and will function as follows: First we train NSL-KDD dataset that has two classes - normal and anomaly, through K2 learning process. It will take the data from the dataset and train it to detect certain patterns, and then decide if it's a normal behaviour or an anomaly. The K2 training will consist of Bayesian Network classifier which will help detecting anomalies in the dataset. After the Bayesian Network detection is over, the system will go through a second training using the Random tree classifier to detect any threats that the Bayesian Network might have missed. Then the dataset will go through a third training using Random forest classifier to detect any anomalies that might have been missed by the previous classifiers. When the training is complete we will open new connection to the junction tree which will connect every node to a parent and predict anomalies from the normal behaviour.

### 6.7. *Overall Detection Accuracy Results*

The testing was initially done on a smaller dataset (20% on NSL-KDD dataset). The accuracy was 99.67% where few instances were classified wrong. The overall results were high with 83 instances classified wrong out of 25109 instances. 29 anomalies were classified normal and 54 normal were classified anomalies. The reason we got lower results from what we tested in Weka is because we used 20% of the dataset, and trees perform better on larger datasets. Refer to table 4. Then we performed the test on the full NSL-KDD dataset. The results for the full dataset was very high, i.e. 99.99%.

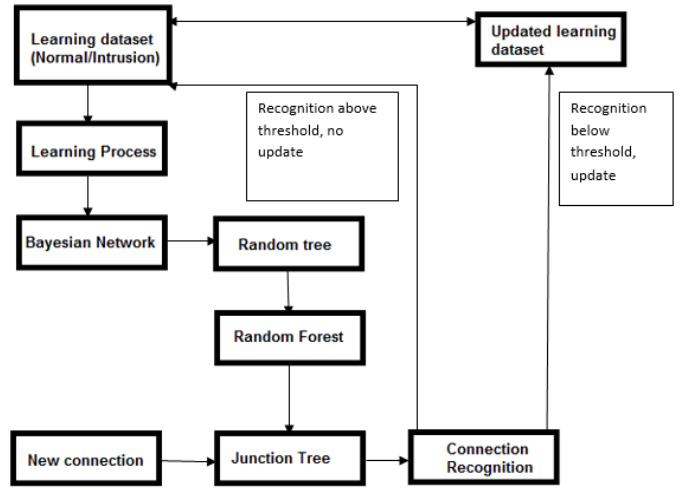

Fig. 6. System Design Classes and their relationship

From the table 4 it is clear that combining the three classifiers gives the highest accuracy with intrusion detection rate of 99.99% with an extremely low false alarm rate of 0.01%. This is quite encouraging compared with all other categories. Although it is slower to build the model with 24.97 seconds than all other classifiers, the classifier makes up for it in high detection rate. Further, high F-value of 100% and high precision 100% and recall 100%, makes it a very good result overall, which is why we chose to combine and use the classifiers

Table 4. Detection Accuracy Comparison

|  | Naïve Bayes | Random Forest | Random Tree | Discretize Filter Naive Bayes | Filter Method Naïve Bayes | Wrapper Method Naïve Bayes | Combined Classifiers 20% NSL-KDD | **Combined Classifiers Full NSL-KDD** |
|---|---|---|---|---|---|---|---|---|
| Detection Rate (%) | 90.38 | 99.89 | 99.77 | 97.13 | 90.72 | 96.22 | 99.67 | **99.99** |
| False Positive Rate (%) | 0.134 | 0.002 | 0.003 | 0.054 | 0.151 | 0.42 | 0.005 | **0.001** |
| Model Building Time (Sec) | 1.57 | 22.33 | 2.59 | 0.12 | 2.19 | 1.62 | 3.23 | **24.97** |
| Precision (%) | 0.89 | 0.999 | 0.998 | 0.954 | 0.88 | 0.964 | 0.996 | **1** |
| Recall (%) | 0.936 | 0.999 | 0.998 | 0.994 | 0.958 | 0.966 | 0.998 | **1** |
| Root Mean Squared Error | 0.3058 | 0.0313 | 0.0479 | 0.1612 | 0.3007 | 0.193 | 0.116 | **0.0086** |

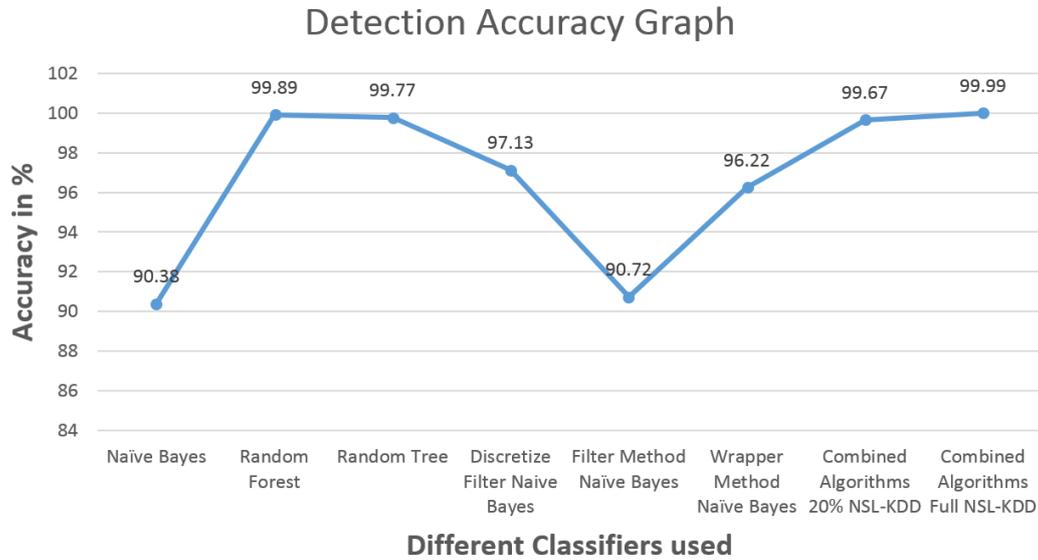

Fig. 7. Detection accuracy comparison graph

instead of using it separately. The accuracy comparison graph can be seen in figure 7.

### 7. Discussion and Limitations of our Work

The work done focuses mainly on the four attack types - DOS, U2R, R2L and Probe in the dataset used. So the attacks outside these could go unnoticed, as we have not trained and tested them. Our focus was to show that a hybrid version of classification algorithms can work better on a given intrusion detection dataset rather than individual ones. The use of an active or passive traffic analyser in conjunction with our software will help to monitor new attacks. So the use of network security monitors like "bro" can only complement our findings. Even though we have used NSL-KDD dataset which is done in 1999, the kind of network attack types remain quite similar even now, even though there are emerging and new kinds of attacks. It is true that some new attacks inside and outside of these categories will always evolve. We are sure that if we train the software with newer attack types, such attacks could as well be detected too, as the software is intelligent and adaptable to changes. The work we have done is only at a prototype level where we have not tested the software with real-time traffic. It may not be that easy to generate a similar dataset with real time traffic with different kinds of attacks as in NSL-KDD dataset as it was generated in an exhaustive manner in a military network. We will try to address this in our future work.

### 8. Conclusion

In this paper we have outlined the importance of intrusion detection systems, and have analyzed the performance of some of the detection classifiers in relation to NSL-KDD dataset. Finally we developed a software system in Java to detect intrusion on networks using the same dataset. Bayesian network has the capabilities to provide auto detection, and they learn from auditing data which can be either normal or abnormal. This was combined with Random tree and Random forest classifiers to get better detection accuracy. The system demonstrated a high performance in detecting intrusion with 99.67% accuracy on 20% of the NSL-KDD dataset and 99.99% accuracy on the full dataset with a model building time of 24 seconds. The higher accuracy was because we used trees in the classifiers, and they tend to give a higher performance when used on large datasets. It should also be noted that for different datasets different individual classifiers may work well or bad, but a combination of best performing classifiers can behave more consistently across different datasets.

### References

1. Hofmann A. and Sick, B. (2011). "Online Intrusion Alert Aggregation with Generative Data Stream Modeling," *Dependable and Secure Computing*, IEEE Transactions on, vol. 8, pp. 282-294.

2. Neethu, B. (2012). "Classification of Intrusion Detection Dataset using machine learning Approaches,"

*International Journal of Electronics and Computer Science Engineering,* vol. 1, pp. 1044-51, 2012.

3. Bace, R. (1999). An Introduction to Intrusion Detection and Assessment: For System and Network Security Management. *ICSA White*, 2, p.32.

4. Bayesnets.com, (2014). *Bayes nets*. [Online] Available at: http://www.bayesnets.com/ [Accessed 25 May. 2014].

5. CitizenNet and Blackwell, A. (2012). A Gentle Introduction to Random Forests, Ensembles, and Performance Metrics in a Commercial System. Accessed online on 21 August 2014. [Online] Available at: http://citizennet.com/blog/2012/11/10/random-forests-ensembles-and-performance-metrics/

6. Cooper, G. and Herskovits, E. (1992). A Bayesian method for the induction of probabilistic networks from data, Machine Learning. 9, pp.309-347.

7. Thomas, V. Sharma and N. Balakrishnan (2008), "Usefulness of DARPA dataset for intrusion detection system evaluation", Proceedings of SPIE 6973, Data Mining, Intrusion Detection, Information Assurance, and Data Networks Security.

8. K. Reddy, M. IAENG, V. N. Reddy, and P. G. Rajulu, (2011). "A Study of Intrusion Detection in Data Mining," *World Congress on Engineering,* vol. III, July 6-8.

9. G. Kalyani and A. J. Lakshmi, (2012). "Performance Assessment of Different Classification Techniques for Intrusion Detection," *IOSR Journal of Computer Engineering (IOSRJCE),* vol. 7, no. 5, pp. 25-29, 2012.

10. G. V. Nadiammai and M. Hemalatha, (2012). "*Perspective analysis of machine learning classifiers for detecting network intrusions*," IEEE Third International Conference on Computing Communication & Networking Technologies (ICCCNT), India, pp. 1-7.

11. IDS, A. (2014). *An Introduction to IDS | Symantec Connect Community*. [Online] Available at: http://www.symantec.com/connect/articles/introduction-ids [Accessed 25 May. 2014].

12. J. McHugh, "Testing intrusion detection systems: A critique of the 1998 and 1999 DARPA intrusion detection system evaluations as performed by Lincoln Laboratory". ACM Transactions on Information and System Security, vol. 3, no. 4, pp. 262–294, 2000.

13. James P. Anderson, (1980). "Computer security threat monitoring and surveillance," *Technical Report,* Fort Washington, Pennsylvania, USA.

14. Jemili, F., Zaghdoud, M. and Ben Ahmed, M. (2007). A framework for an adaptive intrusion detection system using Bayesian network. pp.66--70.

15. KDD Cup 1999 Data (2014), Data and Task description, Online: http://kdd.ics.uci.edu/databases/kddcup99/ (accessed on May 2014).

16. M, Oded, and R, Lior. (2010). Random Trees in the "Data Mining and Knowledge Discovery Handbook", Springer.

17. M. A. Aydin, et al., (2009). "*A hybrid intrusion detection system design for computer network security*," Computers & Electrical Engineering, vol.35, pp. 517-526.

18. M. K. Siddiqui and S. Naahid, (2013), Analysis of KDD CUP 99 Dataset using Clustering based Data Mining, International Journal of Database Theory and Application, 6(5), pp.23-34.

19. M. Tavallaee, E. Bagheri, L. Wei, and A. A. Ghorbani, (2009). "A detailed analysis of the KDD CUP 99 dataset," *in IEEE Symposium on Computational Intelligence for Security and Defense Applications, CISDA* . pp. 1-6.

20. M. Mahoney and P. Chan, "An analysis of the 1999 DARPA/Lincoln Laboratory evaluation data for network anomaly detection". In Recent Advances in Intrusion Detection, vol. 2820 of Lecture Notes in Computer Science, pp. 220–237. Springer Berlin / Heidelberg, 2003.

21. NSL-KDD. (2014). *The NSL-KDD Dataset*. [Online] Available at: http://nsl.cs.unb.ca/NSL-KDD/ [Accessed: 4 Mar 2014]

22. P, A, M., Abraham, A. and Patra, M. R. (2012). A hybrid intelligent approach for network intrusion detection. *Procedia Engineering*, 30 pp. 1--9.

23. P. Srinivasulu, D. Nagaraju, P. R. Kumar, and K. N. Rao, (2009). "Classifying the Network Intrusion Attacks using Data Mining Classification Methods and their Performance Comparison," *IJCSNS* International Journal of Computer Science and Network Security, vol. 9, no.6, pp. 11-18.

24. Pearl, J. (1988). Probabilistic Reasoning in Intelligent Systems. *Morgan Kaufmann*, 0-934613, pp.73-7.

25. R. Lippmann, J. W. Haines, D. J. Fried, J. Korba, and K. Das, (2000). "The 1999 DARPA off-line intrusion detection evaluation," *Computer Networks*, vol. 34, no. 4, pp. 579-595.

26. S. Benferhat and K. Tabia, "*Integrating Anomaly-Based Approach into Bayesian Network Classifiers*," (2009). e-Business and Telecommunications, pp. 127-139.

27. S. Subramanian, V. B. Srinivasan, and C. Ramasa, (2012). "Study on Classification Classifiers for Network Intrusion Systems," pp. 1242-1246.

28. Schultz, M. G., Eskin, E., Zadok, E., and Stolfo, S. J. (2001). "Data Mining Methods for detection of New

Malicious Executables," *IEEE Symposium on Security and Privacy, Columbia University*, pp.38-49.

29. Snort. (2014). The open Source network intrusion detection system [Online]. Available: http://www.snort.org.

30. Stat.berkeley.edu (2014). *Random forests - classification description*. [Online]Available at: http://www.stat.berkeley.edu/~breiman/RandomForests/cc_home.htm [Accessed 25 May 2014]

31. StatSoft (2014). Naive Bayes Classifier. [Online] Available at: http://www.statsoft.com/textbook/naive-bayes-classifier [Accessed 25 August 2014]

32. T .Elvis, et al., (2004). "A serial combination of anomaly and misuse IDSes applied to http traffic", Proceedings of the 20th Annual Computer Security Applications Conference, pp.428-437.

33. T. Hwang, T.Lee, and Y. Lee, (2007). "A Three-tier IDS via Data Mining Approach," *3rd annual ACM workshop on Mining network data*, pp. 1-6.

34. Tavallaee, M., Bagheri, E., Lu, W. and Ghorba ni, A. (2009). A detailed analysis of the KDD CUP 99 dataset. In IEEE Symposium on Computational Intelligence for Security and Defense Applications, Cisda, pp. 1–6.

35. W. Lee and S. J. Stolfo, "Data mining approaches for intrusion detection (1998).*," 7th USENIX Security Symposium,* San Antonio, TX.

36. WEKA. (2014). *Weka 3 - Data Mining with Open Source Machine Learning Software in Java*. [Online] Available at: http://www.cs.waikato.ac.nz/ml/weka/ [Accessed: 4 Mar 2014].